\documentclass[a4paper,fleqn,usenatbib,useAMS]{mnras}
\usepackage{graphicx}	
\usepackage{amssymb}	
\usepackage{multicol}   
\usepackage{bm}		
\usepackage{amsmath}

\title[Images and Spectra of TCAF with outflows]{Images and Spectra of 
Time Dependent Two Component Advective Flow in Presence of Outflows}

\author[A. Chatterjee, S. K. Chakrabarti, H. Ghosh \& S. K. Garain]
{Arka Chatterjee$^{1}$\thanks{E-mail: arka019icsp@gmail.com}, Sandip K. Chakrabarti$^{2,1}$\thanks{E-mail: chakraba@bose.res.in}, 
Himadri Ghosh\thanks{himadri.ghosh@heritageit.edu}$^{3,1}$ \& Sudip K. Garain$^{4}$\thanks{E-mail: sgarain@nd.edu} \\
$^{1}$Indian Centre for Space Physics, Chalantika 43, Garia Station Rd.,
             Kolkata, 700084, India\\
$^{2}$S. N. Bose National Centre for Basic Sciences, Salt Lake,
              Kolkata, 700098, India \\
$^{3}$Heritage Institute of Technology, Kolkata, 700107, India\\
$^{4}$Department of Physics, University of Notre Dame, Notre 
             Dame, IN 46556, USA}
\date{Accepted XXX. Received YYY; in original form ZZZ}

\pubyear{2018}

\begin{document}
\label{firstpage}
\pagerange{\pageref{firstpage}--\pageref{lastpage}}
\maketitle

\begin{abstract}
Two Component Advective Flow (TCAF) successfully explains the spectral and temporal
properties of outbursting or persistent sources. Images of static TCAF 
with Compton cloud or CENtrifugal pressure supported Boundary Layer (CENBOL) 
due to gravitational bending of photons have been studied before. In this paper, 
we study time dependent images of advective flows around a Schwarzschild black 
hole which include cooling effects due to Comptonization of soft photons from a 
Keplerian disks well as the self-consistently produced jets and outflows. We show 
the overall image of the disk-jet system after convolving with a typical beamwidth. 
A long exposure image with time dependent system need not show the black hole horizon 
conspicuously, unless one is looking at a soft state with no jet or the system along 
the jet axis. Assuming these disk-jet configurations are relevant to radio emitting 
systems also, our results would be useful to look for event horizons in high accretion 
rate Supermassive Black Holes in Seyfert galaxies, RL Quasars.

\end{abstract}




\begingroup
\let\clearpage\relax
\endgroup
\newpage

\section{Introduction}
 
Spectral and temporal properties of galactic and extra-galactic black holes 
suggest time dependence of various degrees.  Imaging a black hole accretion disk 
would therefore not be accurate without inclusion of the time dependence of 
various disk components. There is a class of stellar mass black hole 
candidates, known as the outbursting sources, which totally changes its disk configurations
in a matter of few days, when it goes through transitions of its spectral states from 
hard to hard intermediate to soft intermediate and then soft states during their
rising phase. Recently, through fits of data with Two Component Advective Flow (TCAF) solution
as prescribed by Chakrabarti \& Titarchuk (1995; hereafter, CT95), the accretion rates
of the halo and disk components, the sizes of the hot electron cloud (`Compton cloud') etc.
were all shown to be highly variable. The Compton cloud in this solution is the 
inner part of the advective halo, puffed up due to slowing down by the centrifugal barrier,
which not only inverse Comptonizes the soft photons to produce high energy X-rays, but also 
produces outflowing matter just as a normal boundary layer. The disk component on the 
equatorial plane supplies soft photons. In the absence of the CENBOL
as in the soft state, the flow behaves as a standard thin disk with the inner edge 
close to the inner stable circular orbit or ISCO (Shakura \& Sunyaev, 1973) and no jets or outflows 
are seen.  In class-variable sources, such as  GRS 1915+105, temporal variation in 
significant count rates and state transitions such as (burst-on and burst-off states) 
are seen in a matter of few seconds (Muno et al. 1999). This source also exhibits strong 
relativistic radio jets (Fender et al. 1999). Along with the monotonic changes in the CENBOL sizes, the 
frequency of the Quasi-Periodic Oscillations (QPOs) is also seen to change
monotonically (Chakrabarti et al. 2005). Theoretical work which forms the basis of TCAF 
is in Chakrabarti (1996) and references therein. 

The time dependent simulations (Giri, Garain \& Chakrabarti 2013 
and references therein) show the formation of two components in the flow 
with the Keplerian disk on the equatorial plane and advective halo surrounding it
formed the shock and outflows exactly as envisaged in CT95. Outflows are intrinsically associated with 
accretion mechanism (see C90 and Chakrabarti 1996 for further details) in this solution. The explicit 
ratio of mass outflow rate to the inflow rate was first computed in Chakrabarti (1999)
and was found to strongly depend on the compression ratio at the shock.
From Fig. 4 of Giri \& Chakrabarti (2012), it is clear that the ratio of the outflow to 
the inflow rate increases with decreasing viscosity parameter and the inviscid flow 
produces maximum outflows. In outbursts or class variable sources, outflows are found to be 
sporadic and highly variable. Thus, it is very important to have the 
time dependent spectra and images of TCAF where outflows are self-consistently 
generated from the CENBOL itself. Cooling via Comptonization is an essential mechanism by which the 
CENBOL reduces its size making the spectrum softer. While the CENBOL lasts,
rough agreement between the cooling and compressional heating timescales causes it
to oscillate and produce the so-called QPOs (Chakrabarti et al. 2015 and references 
therein). Detailed hydrodynamic simulations in presence of Compton cooling 
are in Ghosh, Garain \& Chakrabarti (2011) and Garain, Ghosh \& Chakrabarti 
(2014) where effects of cooling on the spectra, shock location and the formation of 
QPOs are described. 

Theoretical efforts for imaging of accretion disks started with the pioneering 
work by Luminet (1978) who drew the image of a standard Keplerian  disk around a 
Schwarzschild black hole (Shakura-Sunyaev, 1973). The method of image construction was done from the 
observers end. This technique is very useful as photons were not lost while 
tracking them from the observer. Following this, Fukue \& Yokohama (1988) published 
the colored version of the image. With that, they also studied the occultation
effect on accretion disk light curve. A few years later, Viergutz (1992) generalized
the transfer functions for Kerr geometry. Marck (1996), for the first time, produced 
the Keplerian disk image using the actual Ray-Tracing mechanism. Bromley et al. (2001) 
added polarization determination. Dexter \& Agol (2009) first published `geokerr', a 
public code to compute images in Kerr geometry. `GYOTO' (Vincent et al., 2011),
`YNOGK' (Yang \& Wang, 2013), `GRTRANS' (Dexter, 2016) came into public domains. The
focus of these works was to prepare one to interpret results from the Event 
Horizon Telescope (EHT) whose objective is to identify the horizon of the supermassive 
black hole at the centre of our Galaxy. 

Armitage \& Reynolds (2003), first performed MHD simulation of Keplerian disk and
produced the images where the disk structure remained the same at all the time. 
More recently, Broderick \& Loeb (2009) presented results where simulated images 
of force free jets (Tchekhovskoy et al. 2008) launched from M87 are shown for various 
model parameters and polarization maps of the magnetically driven jets are obtained.
However, under TCAF paradigm, one does not have to introduce magnetic field to generate 
outflows. The hydrodynamic inflow produces time varying outflows self-consistently. Magnetic 
field may be useful for collimation and acceleration purposed further out.

GRMHD simulations in the context of our Galactic centre were mostly performed 
(Ohsuga et al. 2005; Noble et al. 2007; M\`oscibrodzka et al. 2009, 2011; Dexter et al. 2010; 
Hilburn et al. 2010; Dexter \& Fragile 2012; Dolence et al. 2012; Shcherbakov, Penna 
\& McKinney 2012, S\c adowski et al. 2013, Roelofs et al. 2017) considering a very 
low accretion rate as applicable. Cooling term was neglected as appropriate for Sgr A*. 
In fact, Dibi et al. (2012) added the cooling term and found the effect to be negligible in case 
Sgr A*. Drappeau et al. (2013) constrained the mass accretion rate to be $\sim10^{-9}~M_{\odot}/yr$
with a highly spinning central black hole. This accretion rate is much lower 
than that of the earlier result ($\sim10^{-5}~M_{\odot}/yr$) predicted by Coker \& Melia 
(1997). A detailed review on the simulations and observational possibilities of the black hole 
shadow of Sgr A* was presented by Falcke \& Markoff (2013). But, most of the earlier studies were 
based on the RIAF model to simulate spectra and images of Sgr A* only. However, there exists 
a large number of SMBHs in Radio Loud Quasars, Seyfert galaxies, where the accretion rate 
could be substantially high (Bian \& Zhao, 2003) and thus formation of disk and cooling becomes 
important. Another situation which might induce a sudden high mass inflow is from tidal 
disruption of neighboring object (Gillessen et al. 2012).

Chatterjee, Chakrabarti \& Ghosh, (2017a, hereafter CCG17a) showed the static images of TCAF
with general relativistic thick disks acting as the Compton cloud and relativistic 
Keplerian disk (Page \& Thorne, 1974) acting as the source of seed photons. Images for various disk 
rates and inclination angle are shown with their observable spectral counterparts.
The variation of optical depth of the CENBOL medium is achieved by changing the 
halo rate ($\dot{m}_h$). They have shown that by increasing the optical depth, the Keplerian 
disk from the other side is completely blocked by the CENBOL medium. Also, with increasing
inclination angle, the spectral hardening can be seen in their work. Following the
same geometry, Chatterjee, Chakrabarti \& Ghosh, 2017b, (hereafter CCG17b) simulated
the time lags of various energy bins and compare the simulated results with the 
observed counterparts. 

In this paper, for the first time, we present time dependent images of TCAF
with Comptonization and outflows. Basic simulation process by hydrodynamic TVD 
code is explained \S 2. Thermodynamic properties of CENBOL and Keplerian 
disk are presented in \S 3. Section deals with Monte-Carlo simulation of Comptonization 
and cooling mechanism. The spectra, time dependent images with and without 
cooling effects are presented in the results section. Convolved images of accretion
disk at different spectral states are presented. We discuss our results and conclude
this paper in the final section.
              
\section{Time Dependent Simulations of TCAF}
Hydrodynamic simulation procedure is similar to what has been earlier reported in Ryu et al. (1997),
Giri et al. (2010), Giri \& Chakrabarti (2012, 2013), Giri, Garain \& Chakrabarti (2015). 
Governing equations of TVD are explicitly presented by Ryu et al. 1996; Molteni et al. 
1996; Giri \& Chakrabarti 2012. The conservation equation of mass, momentum and energy for an 
axisymmetric, inviscid, non-magnetic flow takes the following form
\begin{equation}
\frac{\partial \boldsymbol {q}}{\partial t} + \frac{1}{r} \frac{\partial\boldsymbol {F_1}}{\partial r} + 
\frac{\partial\boldsymbol {F_2}}{\partial r} + \frac{\partial\boldsymbol {G}}{\partial z} = \boldsymbol {S},
\end{equation}
where
$\boldsymbol{q}= \left (\begin{array}{c}
\rho\\
\rho v_r\\
\rho v_{\theta}\\
\rho v_{z}\\
E
\end{array} \right ) $ is the state vector,  \\
$\boldsymbol{F_1}= \left (\begin{array}{c}
\rho v_{r}\\
\rho v_{r}^{2}\\
\rho v_{r}v_{\theta}\\
\rho v_{r}v_{z}\\
(E+\rho)v_{r}
\end{array} \right ), $
$\boldsymbol{F_2}= \left (\begin{array}{c}
0\\
p\\
0\\
0\\
0
\end{array} \right ), $

$\boldsymbol{F_2}= \left (\begin{array}{c}
\rho v_{z}\\
\rho v_{r}v_{z}\\
\rho v_{\theta}v_{z}\\
\rho v_{z}^2 + p\\
(E+\rho)v_{z}
\end{array} \right )$ are the flux functions.

The source function is defined as\\ 
$\boldsymbol{S}= \left (\begin{array}{c}
0\\
\frac{\rho v_{\theta}^2}{r}-\frac{\rho r}{2(\sqrt{r^2+z^2}-1)^2\sqrt{r^2+z^2}}\\
-\frac{\rho v_{r}v_{\theta}}{r}\\
-\frac{\rho z}{2(\sqrt{r^2+z^2}-1)^2\sqrt{r^2+z^2}}\\
-\frac{\rho (rv_{r}+zv_{z})}{2(\sqrt{r^2+z^2}-1)^2\sqrt{r^2+z^2}}
\end{array} \right ). $

The energy density $E=p/(\gamma -1)+\rho(v_{r}^2+v_{\theta}^2+v_{z}^2)/2$ is the sum 
of thermal and kinetic energy of the infalling matter. The equation corresponding
to energy density can be extracted as
\\

$
\frac{\partial {E}}{\partial t} + \frac{1}{r} \frac{\partial {(E+\rho)v_{r}}}{\partial x} + 
+ \frac{\partial {(E+\rho)v_{z}}}{\partial z} =  
-\frac{\rho (rv_{r}+zv_{z})}{2(\sqrt{r^2+z^2}-1)^2\sqrt{r^2+z^2}}.
$
\\

During cooling, change in $E$ directly modifies the flow configurations following Eqn (1).

The gravitational field, as far as the hydrodynamics is concerned, is assumed to follow the 
Paczy\'nski \& Wiita (1980) pseudo-Newtonian potential which is given by,
\begin{equation} 
\phi(r,Z)=-\frac{GM_{bh}}{r-r_g}
\end{equation}
where $G$ is the universal gravitational constant, $M_{bh}$ is the mass of the black hole kept at 
$10~M_{\odot}$ throughout the simulation and $r_g=\frac{2GM_{bh}}{c^2}=1$ is the Schwarzschild radius
and $r=\sqrt{r^2+Z^2}$. Velocity of light is denoted by $c$. The flow is considered to be adiabatic 
and axisymmetric with the black hole sitting at centre of the cylindrical co-ordinate system 
defined by $(r,\theta,Z)$. In our current simulation, we have ignored viscosity to 
have the strong and hot shock and consequently highest outflow rate. Thus, the specific 
angular momentum $\lambda$ is constant everywhere. Also, from C89 and C90, it is known that for 
a steady state situation the specific energy $\varepsilon$ is also constant. 
Matter is injected at the outer boundary with specific energy $\varepsilon=0.003$ 
(in units of $c^2$) and specific angular momentum $\lambda=1.8$ )in units of $c r_g$.

TVD scheme was originally developed by Harten (1983) to solve the hyperbolic equations present in 
the hydrodynamic conservation equations. This method works on the second order accuracy by 
modifying first order flux function to use the non-oscillatory results in the second order 
solutions. The details of the code were presented in Ryu et al. (1995, 1997) where the code is 
modified to suit black hole accretion.

Our computational domain is defined as: $0<r<100$ and $0<Z<100$ in $r-Z$ plane with a grid 
resolution of $512\times512$. So, each grid size is $\delta r=0.195 r_g$ and $\delta Z=0.195 r_g$.
The black hole absorption boundary is kept at $1.5 r_g$ after which all informations about the
matter as well as photons are lost. Before starting the simulation, entire grid is filled with a low floor density
($10^{-8}$ where the injected density is $1$ in code unit) to avoid any numerical singularities. 
At the outer boundary sub-Keplerian 
matter is allowed to flow through. During the accretion, the matter follows the inviscid hybrid 
accretion mechanism (Chakrabarti 1990) and forms a shock at a specified position.
We assume axisymmetry and perform our hydro-simulation on the first quadrant of the r-Z plane. 
The black hole is located at the origin of the r-Z coordinate system.  
The initial pressure is chosen such that the sound speed as of the interior material is same as 
the incoming matter. Also, inside the computational domain, all the velocity components are set to 
zero initially. The simulation result does not depend on the chosen initial condition as the 
initial matter is washed away by the incoming matter in a few dynamical time. On the right 
radial boundary, we use an inflow boundary condition. Incoming matter enters the computational 
domain through this boundary. Since the density of the incoming matter is chosen to be unity, it 
has $10^{8}$ times higher pressure and it rushes towards the central black hole practically through 
vacuum until it hits the centrifugal barrier. 

Total dynamical time scale for this simulation is $4000$ which in physical units ($\frac{2GM_{BH}}{c^3}\times f_{t}$, 
where $f_{t}$ contains informations of each time) is about $40$s. In the present context, the hydrodynamical 
simulation has been performed for Galactic black hole candidates. However, these results will also hold for 
supermassive or intermediate mass black holes.

\section{Density and Temperature Profile of the Electron Cloud}
CENBOL is a region of hot electron cloud where inverse Comptonization  of soft photons
originated from the Keplerian disk occurs. The electrons in this region are ultra-relativistic. So, $\gamma=4/3$
is a good assumption for this region. Number density (per $cm^3$) and temperature (in the kilo electron Volt) 
of the electrons are supplied for each grid. The pressure and density is related via polytropic equation 
of state $p=K\rho^{\gamma}$, where $p$ is the pressure, $K$ the entropy constant and $\rho$ is the density 
of the flow. From this, we can write $T\propto \rho^{1/3}$, where $T$ is the temperature of the electron 
cloud. In our studies, the maximum number density of electron is $\sim 10^{19}~per~cm^3$ and temperature is 
$\sim 250~keV$. Hydrodynamical simulation changes the configuration of CENBOL region for each time stamp. The
variation of electron cloud geometry with number density and temperature as given in the color bar are presented in 
Figs. 1.

\begin{figure*}
\centering
\vbox{
\includegraphics[height=5.5cm,width=8.0cm]{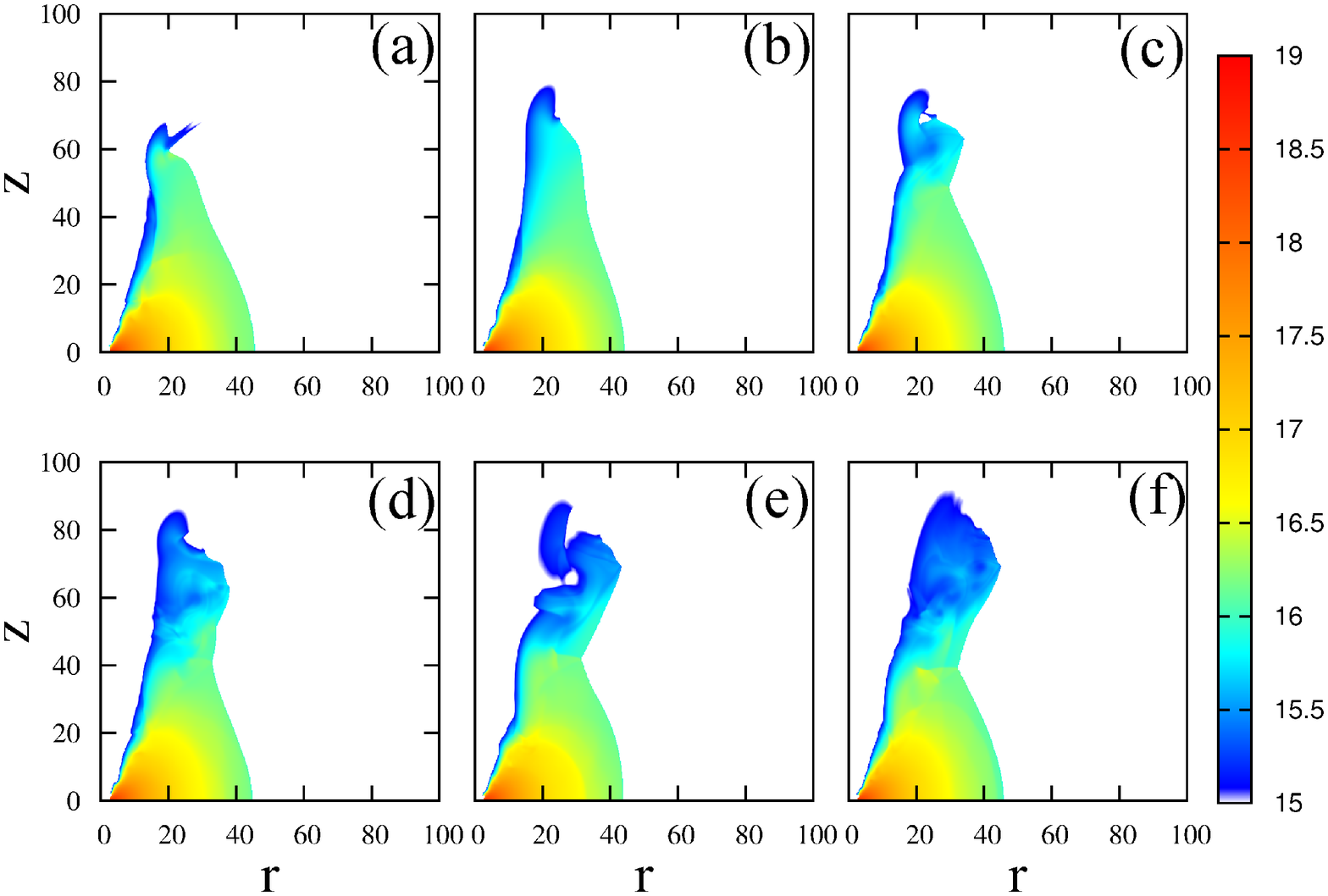}\hskip 0.5cm
\includegraphics[height=5.5cm,width=8.0cm]{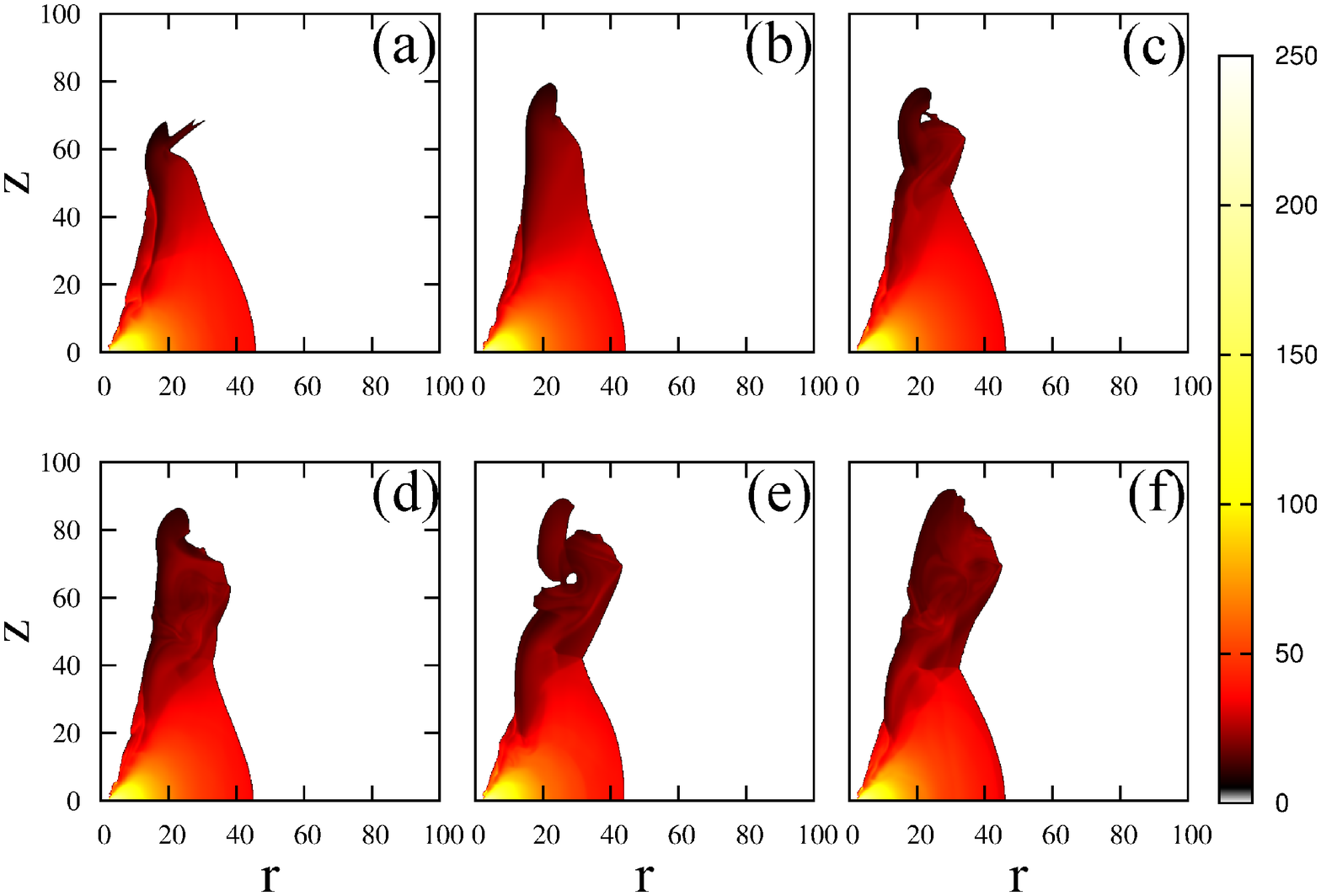}}
\caption{Electron Density (in the units of number per $cm^3$ shown in log scale, left panel) 
and temperature (in the units of keV, right panel) distribution in the post-shock region at 
(a) $t=14.25s$, (b) $t=14.35s$, (c)  $t=14.45s$, (d) $t=14.55$, (e) $t=14.65s$ and (f) $t=14.75s$.}

\end{figure*}

From Fig. 1, the region of outflows can be distinctly seen as a region where temperature and electron 
number density drops rapidly. Over the time, the geometry of the CENBOL region changes as hydrodynamical 
simulation progresses. One can easily differentiate between Fig. 1a and Fig. 1f as the amount of outflow increased 
substantially in the time passed. The inner isobaric contours of the CENBOL region (Figs. 1a-f) 
behave similar to thick disks (Abramowicz et al. 1978; Kozlowski et al. 1978; Paczy\'nski \& Wiita 1980; Begelman, Blandford 
\& Rees 1982; Chakrabarti, 1985a) as pointed out by Molteni, Lanzafame \& Chakrabarti (1994, hereafter MLC94).
 
The TVD method simulates a true accretion mechanism. The injected matter washes away the
initial low floor matter and attains a quasi steady state. Inflowing matter undergoes a shock due 
to the centrifugal barrier. After the shock, matter starts to flow towards black hole. The mass 
absorption rate through the black hole boundary ($1.5~r_g$) is represented in the Fig. 2. It can
be seen that the mass absorption rate after the transient phase saturates at $\sim75\%$ of the 
injected mass.   

\begin{figure}
\vskip 0.4cm
\centering
\vbox{
\includegraphics[width=1.0\columnwidth]{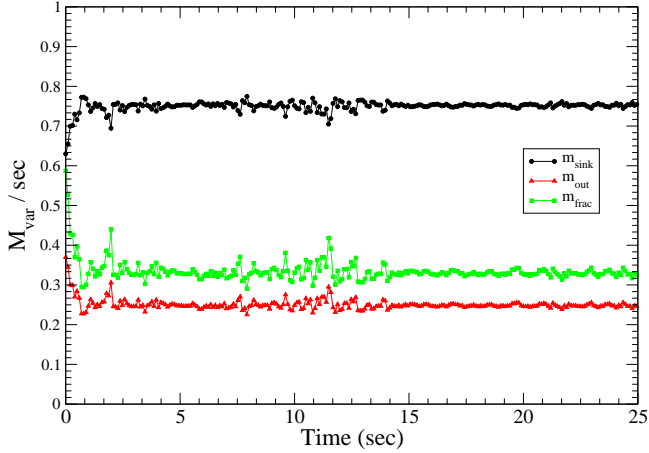}}
\caption{Fraction of mass absorption (circle-black), outflows (triangle-red) and the ratio of outflowing
matter to that of absorption (square-green) through the inner boundary with time are shown. Injected 
halo rate ($\dot{m}_h$) is kept fixed at $0.1$.}
\end{figure}

First simulation of this kind was performed by MLC94 followed by Molteni, Ryu \& Chakrabarti 1996 (MRC96), Giri \& 
Chakrabarti 2012, 2013 (GC12, GC13). The outflows are generated out of 
inflowing sub-Keplerian matter forming a shock due to centrifugal barrier. 
The oscillation of the shock front after attaining a steady state and the consequence 
of that on light curves are well studied in MLC94 and Giri et al. (2010) and Garain et al. (2013). 
This formation of shock induces a temperature rise in the post-shock region which acts 
as the base of the outflows or Jets. Due to the increase of temperature in the post shock 
region, force due to gradient of pressure pushes matter outward along the vertical 
direction. This creates the outflows. In HD simulations, this is not well collimated. 
One may require magnetic effects for collimation (Chakrabarti, 2013).  
To show details, we have shown the post-shock region only. The 
sub-Keplerian flow of the pre-shock region has much lower density and optical depth as
compared to the post-shock region (GC13). Maximum number of scatterings occur in the 
post-shock region and for an observer trying to capture the image of an accretion disk
in presence of the outflows, the last scattering surface of the CENBOL will appear as 
the shape of the Compton cloud.  

\subsection{Keplerian Disk Acting as the Soft Photon Source}

In order not to include a time dependence in the Keplerian disk, we put it on the equatorial plane, whose
purpose is to supply soft photons as per Page \& Thorne (1974) prescription. We employ the Monte-Carlo simulations
to compute the resulting spectrum which includes the original source photons from this Keplerian disk as well as those
scattered from the CENBOL and the outflows. The Keplerian disk is truncated at the inner edge at the CENBOL surface
and the outer edge is extended up to $100~r_g$ for simulation purpose. 

\begin{equation}
\begin{aligned}
F(r) = 
\frac{F_c(\dot{m}_d)}{(r-3/2)r^{5/2}}  \\                               
\times \bigg[\sqrt{r}-\sqrt{3/2}+
\frac{\sqrt{3/2}}{2}log\bigg(\frac{(\sqrt{r}+\sqrt{3/2})(\sqrt{3}-\sqrt{3/2})}
{(\sqrt{r}-\sqrt{3/2})(\sqrt{3}+\sqrt{3/2})}\bigg)\bigg]\\ 
 \\ and \\                                  
T(r)= \bigg(\frac{F(r)}{\sigma}\bigg)^{1/4}
\end{aligned}
\end{equation}
where, $F_c(\dot{m}_d)=\frac{3m\dot{m}_d}{8\pi r_{g}^3}$, $\dot{m}_d$ is 
the Keplerian disk accretion rate in Eddington unit, $\sigma=\frac{2\pi^5k_{B}^4}{15h^3c^3}$ 
is the Stefan-Boltzmann constant.

Photon flux emitted from the Keplerian disk surface of radius  $r$ to $r+\delta r$ is written as,
\begin{equation}
n_{\gamma}(r)=\bigg[\frac{4\pi}{c^2} \bigg(\frac{k_bT(r)}{h}\bigg)^3 \times \zeta (3)\bigg]~cm^{-2}s^{-1},
\end{equation}
where, $\zeta (3)=\sum^\infty_1{l}^{-3} = 1.202$ is the Riemann zeta function.
So, the rate of photons emitted from the radius $r$ to $r+\delta r$ is given by,
\begin{equation}
dN(r) = 4\pi r\delta rn_{\gamma}(r),
\end{equation}
where the Keplerian disk height $H(r)$ is considered as $0.1$ 
(since $\frac{H(r)}{r} << 1$ for Keplerian disk region).

The disk is divided into several annuli of radial width $D(r) = 0.1$  and of mean 
temperature $T(r)$. The exact number of photons come to be around $\sim 10^{39}$ --$10^{40}$ per 
second for a disk rate of $\dot{m}_d=0.1$. With increasing disk rate this number increases 
rapidly. Using a weightage factor $f_W=\frac{dN(r)}{N_{comp(r)}}$ where $N_{comp}(r) = 10^9$,
we bundle these photons to save computational time.

Soft photon energy is calculated using Planck distribution law for $T(r)$. 
Photon number density ($n_\gamma(E)$) which corresponds to an energy $E$ is given by, 
\begin{equation}
n_\gamma(E) = \frac{1}{2 \zeta(3)} b^{3} E^{2}(e^{bE} -1 )^{-1}.
\end{equation}
The process is similar to the earlier works presented by Ghosh, Chakrabarti \& 
Laurent (2009); Chatterjee, Chakrabarti \& Ghosh (2017a, 2017b).

\section{Monte-Carlo Simulation of Comptonization}

In Monte-Carlo process, the photons inside the CENBOL region are assumed to follow straight line trajectories
in between two scatterings. This enhances the Computational efficiency without sacrificing science. This has been pointed out by 
Laurent \& Titarchuk (1999). The Keplerian disk emits seed photons with six random number associated 
with position and velocity. In reality, the Keplerian disk should have the maximum flux along  
Z-axis and minimum along the equatorial plane. This has been implemented by assuming 
$F_{\nu}=\int I_{\nu}\mathrm{cos}\theta d\Omega$. Inside the CENBOL region,
a critical optical depth ($\tau_c$) is set up using random number corresponding to each 
scattering. The next scattering would occur if the optical depth of a particular photon crosses $\tau_c$.

We choose Klein-Nishina scattering cross section $\sigma$ which is given by:
\begin{equation}
\sigma = \frac{2\pi r_{e}^{2}}{x}\left[ \left( 1 - \frac{4}{x} - \frac{8}{x^2} \right) 
ln\left( 1 + x \right) + \frac{1}{2} + \frac{8}{x} - \frac{1}{2\left( 1 + x \right)^2} \right],
\end{equation}
where, $x$ is given by,
\begin{equation}
x = \frac{2E}{m c^2} \gamma \left(1 - \mu \frac{v}{c} \right),
\end{equation}
$r_{e} = e^2/mc^2$ is the classical electron radius and $m$ is the mass of the electron. This yields
Thomson scattering cross section ($\sigma_{T}$) in low frequencies and Compton scattering cross 
section ($\sigma_{C}$) in higher frequencies like X-rays and $\gamma$-Rays. Between each pair of
scatterings, the gravitational red-shift modifies the energy of the photon. This process
continues until it leaves the CENBOL region or get sucked by the black hole. The process
is similar to that presented in GCL09; Ghosh, Garain, Chakrabarti, Laurent, 2010; Ghosh, Garain, 
Giri, Chakrabarti, 2011 (hereafter GGGC11); CCG17a, CCG17b.   

\begin{figure}
\vskip 0.7cm
\centering
\vbox{
\includegraphics[width=1.0\columnwidth]{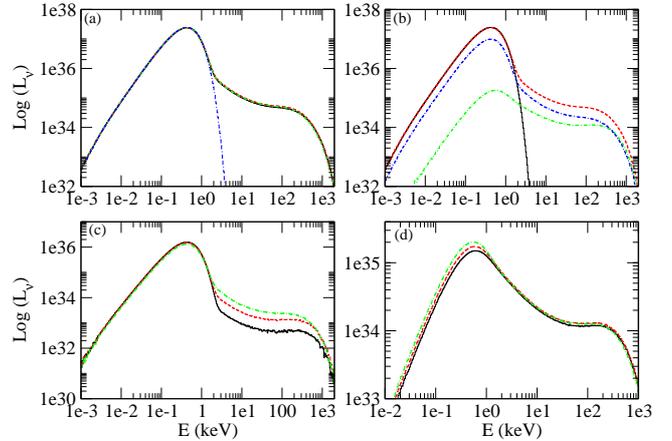}}
\caption{Spectra obtained after Comptonization. (a) The spectra at $t=14.25s$ (Black-Solid), 
$t=14.45s$ (Red-Dotted) and $t=14.75s$ (Green-Dot-Dashed)~and~ injected (Blue-Dot-Dash-Dash) spectrum.
(b) Injected spectrum (Black-Solid), composite spectrum (Red-Dashed), Reflected Spectrum 
(Blue-Dot-Dash-Dash), Outflow spectrum (Green-Dot-Dashed). (c) Inclination dependence
of reflected spectra. The inclination bins are $0^{\circ}-20^{\circ}$ (Black-Solid), $40^{\circ}-60^{\circ}$ 
(Red-Dotted), $70^{\circ}-90^{\circ}$ (Green-Dot-Dashed). (d) Outflow spectra at $t=14.25s$ 
(Black-Solid), $t=14.45s$ (Red-Dotted) and  $t=14.75s$ (Green-Dot-Dashed). Note that the outflow spectra
changed significantly due to the increase in the outflow rate.
Disk accretion rate $\dot{m}_d=0.1$ is kept fixed throughout the simulations.}
\end{figure}

After Comptonization, the spectra at various times are presented in Fig. 3.
Fig. 3a shows the Comptonized spectra of TCAF at $t=14.25s$, $t=14.45s$~and~$t=14.75s$.
The temporal variation in the thermodynamic variables of TCAF has been shown 
in Fig. 1. In the absence of cooling, the Compton cloud remains hot.
Thus, the variations in spectra are very small. The Blue-dot-dot-dashed curve 
in Fig. 3a represents the injected spectra. In (b), various components such as the net spectrum (red-dashed),
injected spectrum (solid-black), reflected spectrum  caused by rescattering of Comptonized radiation by the Keplerian 
disk before it reaches the observer (blue-dot-dot-dashed) and outflow 
spectrum (green-dot-dashed) are shown at $t=14.45s$. It can be clearly seen 
from Fig. 3b that the reflected spectrum contributes more to the net spectrum than 
the outflows spectrum. Fig. 3c shows the reflected spectrum at three inclinations
($0^\circ-22.5^\circ$ (black-solid), $45^\circ-67.5^\circ$ (red-dashed) and $67.5^\circ-
90^\circ$ (green-dot-dashed)) at time $t=14.45s$. The hard energy contribution in 
reflected spectra increases with inclination angles while the soft energy contribution 
decreases. This is an expected result for the reflected component as the reflection dominates 
at high inclinations. With the 
increase of the outflow rate, the spectral contribution of 
the outflow also increases. In Fig. 3d, the outflow spectra have been plotted for 
$t=14.25s$~(black-solid), $t=14.45s$~(red-dotted) and~$t=14.75s$~(green-dot-dashed). 

\subsection{Cooling Process}
During Comptonization, the photons either gain or lose energy. In the context of accretion physics,
the electrons in the CENBOL region is at much higher temperature than that of the photons coming from a Keplerian disk.
So, mostly inverse Compton scattering dominates and the signature of that is prominent in the
emergent spectra. We supply a steady state hydrodynamical configuration obtained from TVD code
where the density and temperature of the cloud is defined in each grid. After the scattering,
the electron from a particular grid may loose or gain $\Delta E$ amount of energy which is
transferred to the photon via Compton mechanism. In a particular state, after all the 
scatterings by the injected photons, the energy of the entire grid is modified to generate a new
hydrodynamical state where the lose or gain of energy from each grid has been accounted. By this
process the final temperature of the cloud steadily decreases. The final temperature of the
electron cloud in a particular grid is expressed as, 
\begin{equation} 
k_BT_{new}(ir,iz) = k_BT_{old}(ir,iz) - \frac{\Delta E}{3dN_e(ir,iz)},
\end{equation} 
where $T_{new}$ and $T_{old}$ are updated and old temperature of electron cloud. $dN_e(ir,iz)$ is 
the number density of the electron in a particular grid denoted by $ir$ and $iz$ and $k_B$ is 
the Boltzmann constant. Details of this process of cooling is earlier presented in Ghosh, Garain, 
Giri \& Chakrabarti 2011 (GGGC11); Garain, Ghosh \& Chakrabarti 2012 (GGC12) and Garain, Ghosh 
\& Chakrabarti 2014 (GGC14). 

\begin{figure}
\centering
\vbox{
\includegraphics[width=1.0\columnwidth]{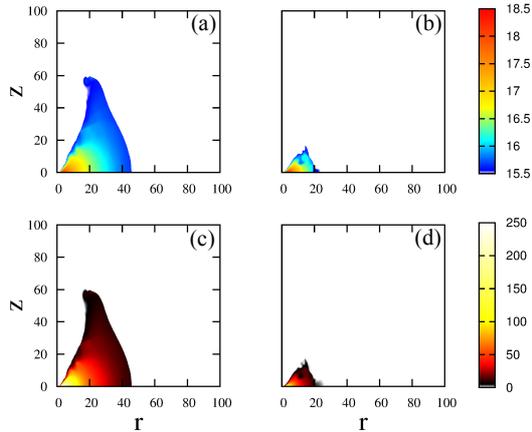}}
\caption{(a-c) Electron number density (in per $cm^3$, upper panel), and (b-d) temperature (in the units of keV, 
lower panel) distribution in the post-shock region at $t=14.25s$. Compton cooling collapses the post-shock 
region (a,c): initial stage of cooling, (b,d): after the cooling process is completed. The shock moves 
inward due to loss of post-shock pressure. Disk rates $\dot{m}_d=0.1$ and $\dot{m}_h=0.1$ are kept constant 
throughout.}
\end{figure}

In Fig. 4, we see that the shock location moves inward as the Compton cooling
starts to affect the post-shock region. Panels (a) and (d) show the CENBOL region
which has almost the same electron density and temperature as Fig. 1 (a). We have
considered this particular data file to see the effect of Compton cooling. Fig 4(a) is presented
when cooling just started to work. Then, after a few iterations cooling starts to reduce the 
CENBOL size. Spectra softens as the size of hotter Compton cloud becomes smaller. Finally, the 
Compton cloud reduces more which makes the accretion disk spectra to be in soft state. 
The corresponding spectra of Panel (b) and (d) suggest the CENBOL region is 
so small that a very few numbers of soft photons are intercepted by the cloud. Thus, the spectrum 
softens and spectral slope increases. The effect of cooling on the outflows are
also visible from Fig. 4. With the inclusion of cooling, the amount of matter outflow rate 
($R_{\dot{m}}$) reduces drastically.

\begin{figure}
\vskip 0.7cm
\centering
\vbox{
\includegraphics[width=1.0\columnwidth]{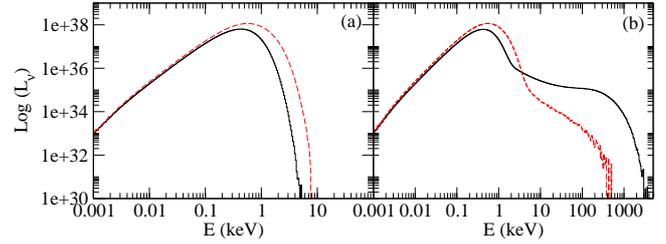}}
\caption{Variation of spectrum due to Compton cooling. Panel (a) corresponds 
to the injected soft spectrum where solid-black is the same spectrum as injected 
to the without cooling cases. Dotted-red corresponds to the injected spectra at final 
stage (Fig. 4b,d). The corresponding emergent Comptonized spectra are presented 
in Panel (b) where solid-black (same as Fig. 3) and dotted-red (for CENBOL represented 
in Fig. 4b,d) curves showing the spectral softening due to Compton cooling.}
\end{figure}

Fig. 6 represents the outflows spectra for two different CENBOLs. Black-solid one is same as
the spectra presented for $t=14.25s$ in Fig. 3d. This is the input TVD file on which Compton
cooling started to act. The CENBOL reduces the its size due to cooling. And outflows component
vanishes with cooling. The outflow spectra shrinks more and more and the it softens. Basically, all
photons in the outflows spectra for the to particular case comes from the base of the jet which is
CENBOL itself.  In this paper, our goal is to show the variation of images and spectra when 
cooling is included. A very detailed study of the effect of cooling on outflows has already 
been reported in GGC12.

\begin{figure}
\centering
\vbox{
\includegraphics[width=1.0\columnwidth]{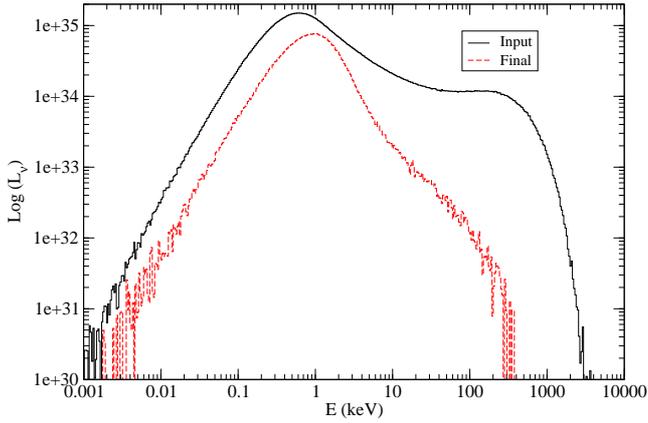}}
\caption{Emergent outflow spectra for corresponding density and temperature 
distribution shown in Fig. 4. Black-Solid curve shows the outflows spectrum 
directly from the output of the TVD code (same as the black curve shown in 
Fig. 3d). Red-dashed represents outflow spectra of the TCAF after final 
stage of cooling.
}
\end{figure}

\section{Ray-Tracing Process}
In Schwarzschild geometry, the non vanishing components Christoffel symbol yield geodesic 
equations from the field equation,
\noindent
\begin{equation}
\frac{d^2x^{\mu}}{dp^2}+ {\Gamma}_{\nu\lambda}^{\mu}\frac{dx^{\nu}}{dp}\frac{dx^{\lambda}}{dp} = 0,
\end{equation}
where ${\mu} = [0,1,2,3]$; $x^{0} = t$, $x^{1} = r$, $x^{2} = \theta$ and $x^{3} = \phi$,
$p$ is the Affine parameter. The four coupled, second order differential equations for photons 
can be reduced to three using energy $P_t=E=(1-\frac{1}{r})\frac{dt}{dp}$ and
angular momentum $P_{\phi}=L=r^{2}\mathrm{sin}^{2}\theta\frac{d\phi}{dp}$ definitions 
(Chandrasekhar, 1985). So, three geodesic equations which dictate the trajectory of photons
can be written as
\begin{equation}
\begin{aligned}
\frac{d^2r}{dt^2} + \frac{3}{2r(r-1)}\bigg(\frac{dr}{dt}\bigg)^2 - \\ (r-1)\bigg(\frac{d\theta}{dt}\bigg)^2 -
 (r-1)r\mathrm{sin}^2{\theta}\bigg(\frac{d\phi}{dt}\bigg)^2 
+ \frac{r-1}{2r^3} = 0,\\
\frac{d^2\theta}{dt^2} + \frac{2r-3}{r(r-1)}\bigg(\frac{d\theta}{dt}\bigg)\bigg(\frac{dr}{dt}\bigg) 
- \mathrm{sin}{\theta}\mathrm{cos}{\theta}\bigg(\frac{d\phi}{dt}\bigg)^2 = 0 ~\mathrm{and}\\
\frac{d^2\phi}{dt^2} +\frac{2r-3}{r(r-1)}\bigg(\frac{d\theta}{dt}\bigg)\bigg(\frac{dr}{dt}\bigg)
 + 2\mathrm{cot}{\theta}\bigg(\frac{d\theta}{dt}\bigg)\bigg(\frac{d\phi}{dt}\bigg) = 0.
\end{aligned}
\label{eq:xdef}
\end{equation}

Velocity components are derived using Tetrad formalism (Park 2006) and are given by,
\begin{equation}
\begin{aligned}
v^{\hat{r}}=\frac{d\hat{r}}{dt}=\frac{r}{(r-1)}\frac{dr}{dt},
~v^{\hat{\theta}}=\frac{d\hat{\theta}}{dt}=\frac{r\sqrt{r}}{\sqrt{(r-1)}}\frac{d\theta}{dt}\\ 
\mathrm{and} ~
v^{\hat{\phi}}=\frac{d\hat{\phi}}{dt}=\frac{r\sqrt{r}\mathrm{sin}{\theta}}{\sqrt{(r-1)}}\frac{d\theta}{dt}.
\end{aligned}
\label{eq:xdef}
\end{equation}

The redshift factor added to connect the source and the observer frame is given by the relation  
\begin{equation}
1+z= \frac{E_{em}}{E_{obs}}=\frac{(P_{\alpha}u^{\alpha})^{em}}{(P_{\alpha}u^{\alpha})^{obs}},
\end{equation}
where, $E_{em}$ and $E_{obs}$ are the energy of emitted and observed photons respectively. 
The observed and emitted fluxes are related via the fourth power of redshift factor 
\begin{equation}
F_{k}^{obs}= \frac{F_{k}^{disk}}{(1+z)^4}.
\end{equation}
This relationship between the observed flux and the observed temperature can be expressed as,
\begin{equation}
T_{k}^{obs} = \bigg(\frac{F_{k}^{obs}}{\sigma}\bigg)^{1/4}.
\end{equation}
Doppler boosting is added during the emitter-observer frame transformation.
The Ray-Tracing process and formation of images are done in the same way as 
as reported earlier in CCG17a, CCG17b.  

\section{Results}

\subsection{Time Dependent Images without Cooling}
The static images of TCAF in presence of Comptonization were presented in CCG17a. The Compton
cloud was modelled using natural angular momentum description of General Relativistic thick disks given by 
Chakrabarti (1985a). The variations of images and spectra were presented for various inclination angles and
disk accretion rates. They show that due to steep variation of density and temperature inside the CENBOL,
the image does not have sharp edges. On the top of that, for an increasing optical depth of the CENBOL region, the photons from 
the Keplerian disk region located opposite to the observer are blocked. In the present case, where TVD code 
was used to create the CENBOL and the outflow self-consistently, the optical depth of the CENBOL medium allows
a fraction of photons from the Keplerian disk to pass through the Compton cloud without getting scattered.
This can be seen in each panels presented in Fig. 7 \& Fig. 8.       

\begin{figure}
\centering
\vbox{
\includegraphics[width=1.00\columnwidth]{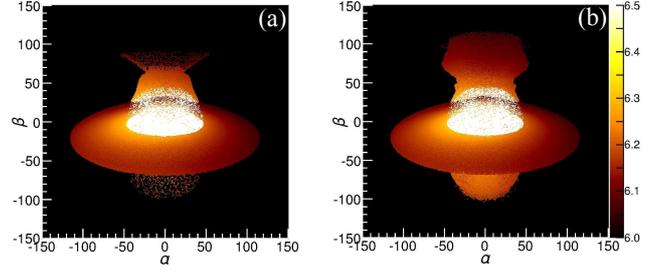}}
\caption{Images of TCAF (in $log(T_{obs})$ scale) seen from an inclination angle of $70^\circ$
in presence of outflows at time $t=14.25s$ ~and~$t=14.75s$. Colorbar is in the range 
$10^{6.0}-10^{6.5}$ Kelvin.}
\end{figure}

Figure 7 is drawn at (a) $t=14.25s$~and~ (b) $t=14.75s$. In panel (a), the amount of outflow is much less as
compared to the panel (b). The photons emitted by the base of the lower jet are mostly blocked by the optically 
thick Keplerian disk. In Fig. 7a, a very low number of photons is visible from the lower 
jet. Due to reflection symmetry of the outflows, these photons are also visible in the upper part of 
the outflows. However, due to Lorentz boosting in the upper jet, its observed temperature
is enhanced and it is overall brighter. The sonic surface of the outflow is the cap-like curved 
bright surface. The flow is subsonic and hotter below this surface.
We include the photons in this plot which are emitted from the regions with optical depth 
greater than unity. This creates a special shape for the base of the outflow.

\begin{figure*}
\centering
\vbox{
\includegraphics[width=2.0\columnwidth]{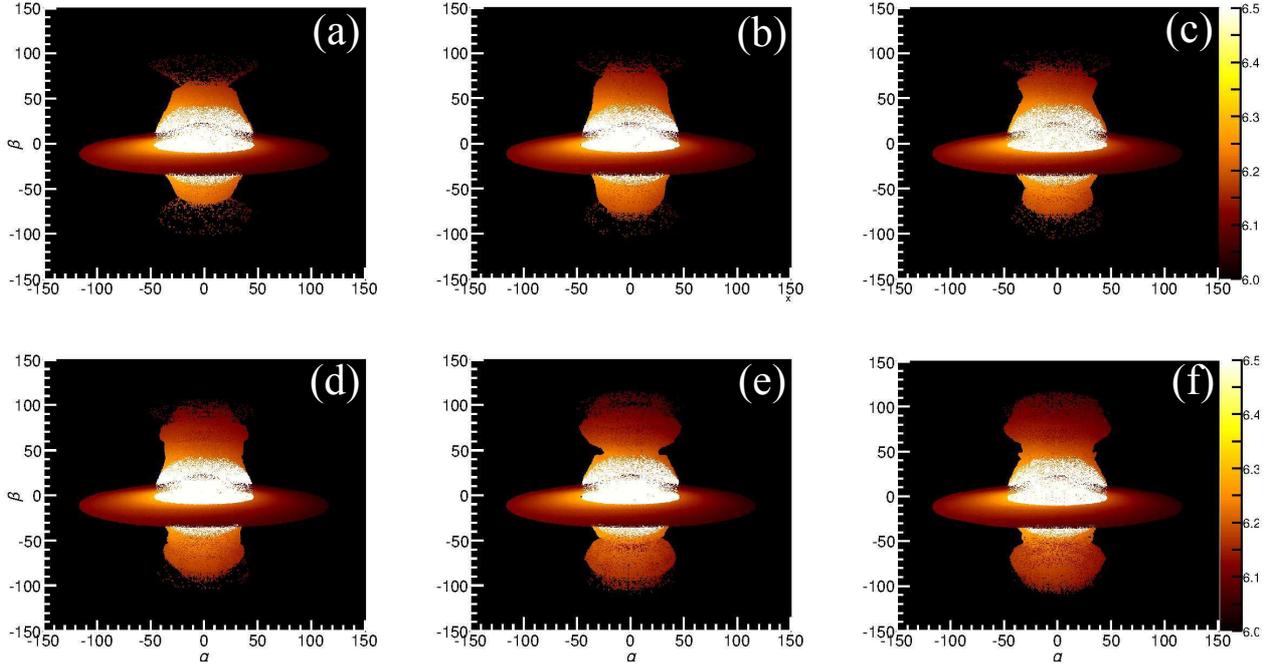}}
\caption{Images of TCAF (in $log(T_{obs})$ scale) in presence of outflows at $t=14.25s$, 
$t=14.35s$, $t=14.45s$, $t=14.55s$, $t=14.65s$~and~$t=14.75s$. Colorbar is in the range $10^{6.0}-
10^{6.5}$ Kelvin and inclination of the observer is at $80^\circ$. Asymmetry in the shape 
is due to difference in Doppler shifts in upper and lower jet components.}
\end{figure*}

Figure 8 shows the dynamical evolution of outflows at $t=14.25s$, $t=14.35s$, $t=14.45s$, 
$t=14.55s$, $t=14.65s$~and~$t=14.75s$. The mass outflow increases with time in this case. The images 
with gravitationally bent rays have been drawn from the perspective of an observer placed at an 
inclination angle of $80^\circ$. The asymmetry of outflows become prominent in this Figure also. The 
energy range is chosen in such a way that the observer temperature variation of the Keplerian 
disk can be found easily. 

\begin{figure}
\centering
\vbox{
\includegraphics[width=1.0\columnwidth]{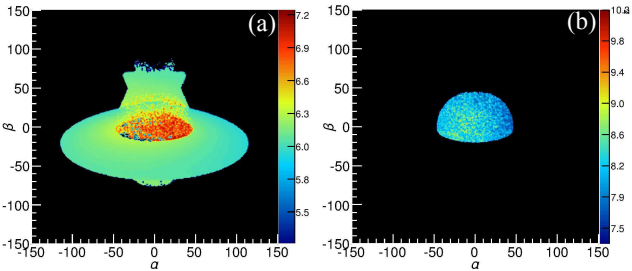}}
\caption{Grid averaged images of TCAF (in $log(T_{obs})$ scale) in presence of outflows at time $t=14.65s$. 
Colorbar is in the range $10^{5.4}-10^{6.7}$ Kelvin for panel (a) and $10^{6.7}-10^{10.3}$ Kelvin for panel
(b). The observer is placed at $70^\circ$. For higher energy range detectors, the Keplerian disk and 
upper part of jet becomes completely invisible to an observer.}
\end{figure}

Figure 9 shows energy dependent images of TCAF. With a higher energy detector which 
works best in the energy range of $1.0-100$ keV, the Keplerian disk might not be seen for disk 
rates lower than $0.1$. With increasing $\dot{m}_d$, the disk may reappear. So, for 
an outbursting source, whose accretion rates change on a daily basis, the images will change for 
any particular energy range of detector. These pictorial changes will be due to the physical 
variations of accretion disk geometry which will be triggered by the thermodynamical properties 
(see, CT95 for details) of the disk.   

In Panel (b), we see the top part of the outflows is also missing in 
a high energy observation. This is due to the velocity profile of the outflowing electrons.
Most of the photons suffer downscattering during their collision in the outflowing electrons. 
In Panel 3d, we see that the Jet spectra extends upto $1000$ keV. But, in reality, 
in our model of non-magnetic jets, maximum hard photons in the Jet spectra are contributed 
only from the base of the jet. The photons that are coming from below the equatorial plane are mostly 
reflected or absorbed by optically thick Keplerian
disk. So, up to $70^\circ$, visibility of the base of a lower jet is screened by the Keplerian
disk itself. But, at $80^\circ$, as in Fig. 8, both sides become visible to the 
observer.   

\subsection{Observed Spectra}
The source spectrum  as presented in the Fig. 3a has a peak 
at $0.6$ keV. This has been modified due to photon bending and shifted to a lower value for 
an observer at higher inclination angle. In the observed spectrum of Keplerian disk has a 
peak at $0.3$ keV for $80^\circ$ inclination angle (Fig. 10). The
corresponding characteristic temperature ranges from $10^{6.0}-10^{6.6}$ Kelvin and this 
is visible in Fig. 9a for $70^\circ$. The observed spectral variation of the black body
part with an inclination angle remains almost the same as reported in CCG17a.  

\begin{figure}
\centering
\vbox{
\includegraphics[width=0.8\columnwidth]{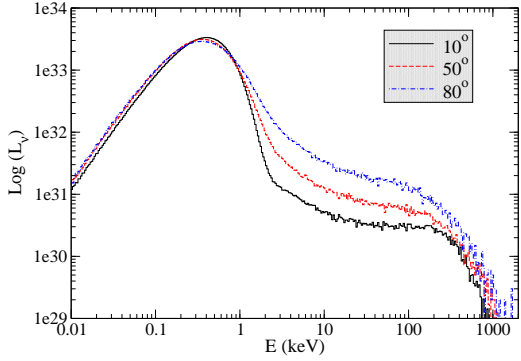}}
\caption{Observed spectra of TCAF seen from inclination angles $10^\circ$, $50^\circ$ 
and $80^\circ$ in presence of Outflows. Increase in the hard photon contribution with inclination 
angle is visible. However, with the inclusion of outflows, the spectral slope increases with 
increasing the inclination angle.}
\end{figure}

From Fig. 10, we see that the contribution of hard photons increases with inclination 
angle. But, the spectral slope is less for low inclination angle. This is to be contrasted 
with our earlier studies (see CCG17a) where the outflow component has not been added. 
Without the outflows, the hard photons in powerlaw component increases and the slope 
of the spectrum decreases with increasing inclination angle. With the inclusion of 
outflows, the spectral slope increases with increasing inclination angle. But, the 
contribution of hard photons increases  more significantly when the observer moves to 
a higher inclination angle.  

\subsection{Images with Cooling}
To demonstrate the effects of Compton cooling, we consider first  the 
data file from hydrodynamic simulations at $t=14.25s$. The electron number density and 
temperature profile of the post-shock region after Compton cooling is presented 
in Fig. 4. Corresponding spectrum in Fig. 5 shows the spectral softening due 
to cooling. From Fig. 11, we see that the size of the CENBOL region is 
substantially reduced due to cooling effect and the optical depth of CENBOL 
region is so low that very few soft photons from Keplerian disk are intercepted
in the Compton cloud. This is the reason of spectral softening. With the inclusion of 
Compton cooling the outflow drastically reduces its size. This effect can also be visible 
in the images of TCAF. As the CENBOL size is reduced, the inner edge of the truncated Keplerian 
disk moves radially inwards. 

\begin{figure}
\centering
\vbox{
\includegraphics[width=0.45\columnwidth]{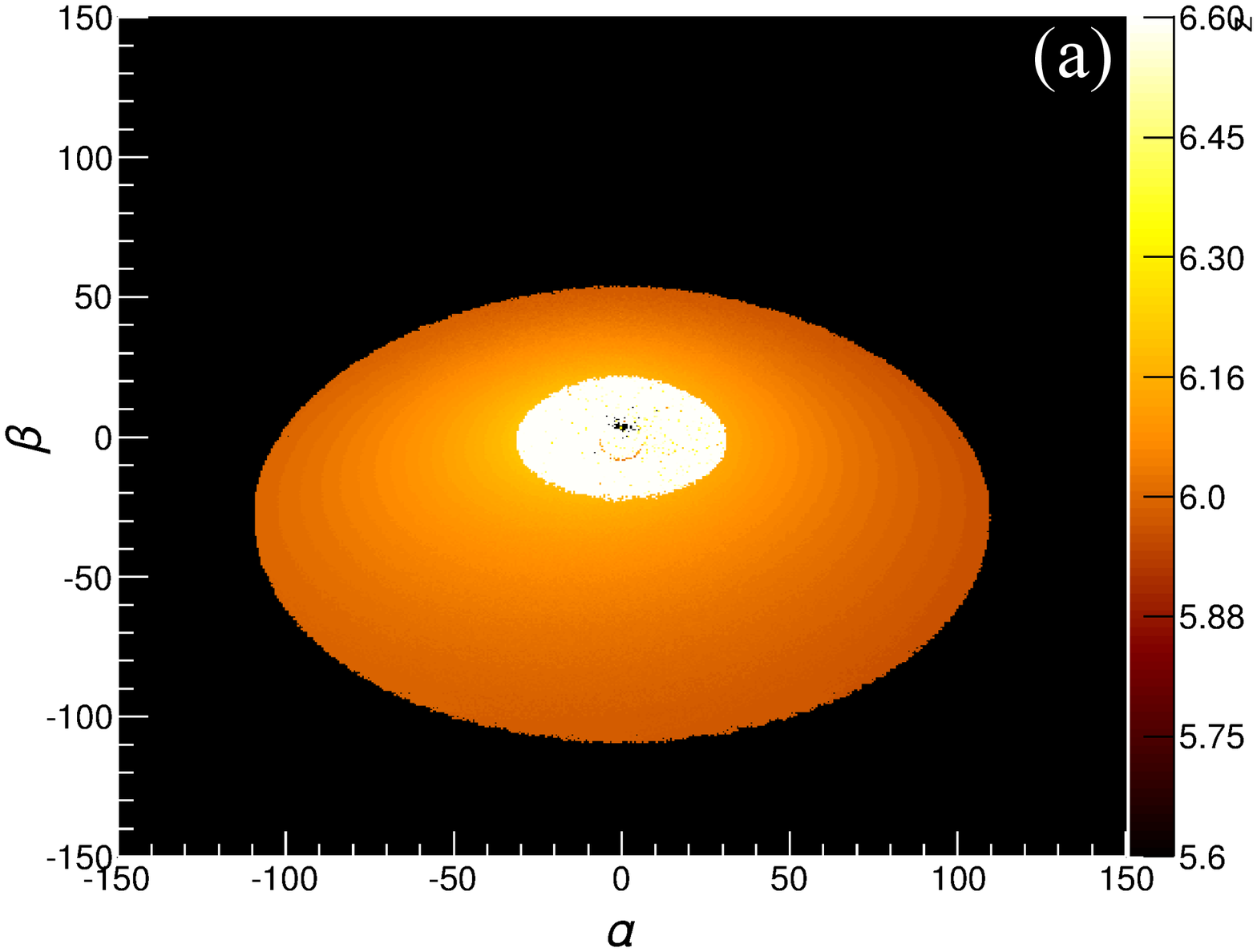}
\includegraphics[width=0.45\columnwidth]{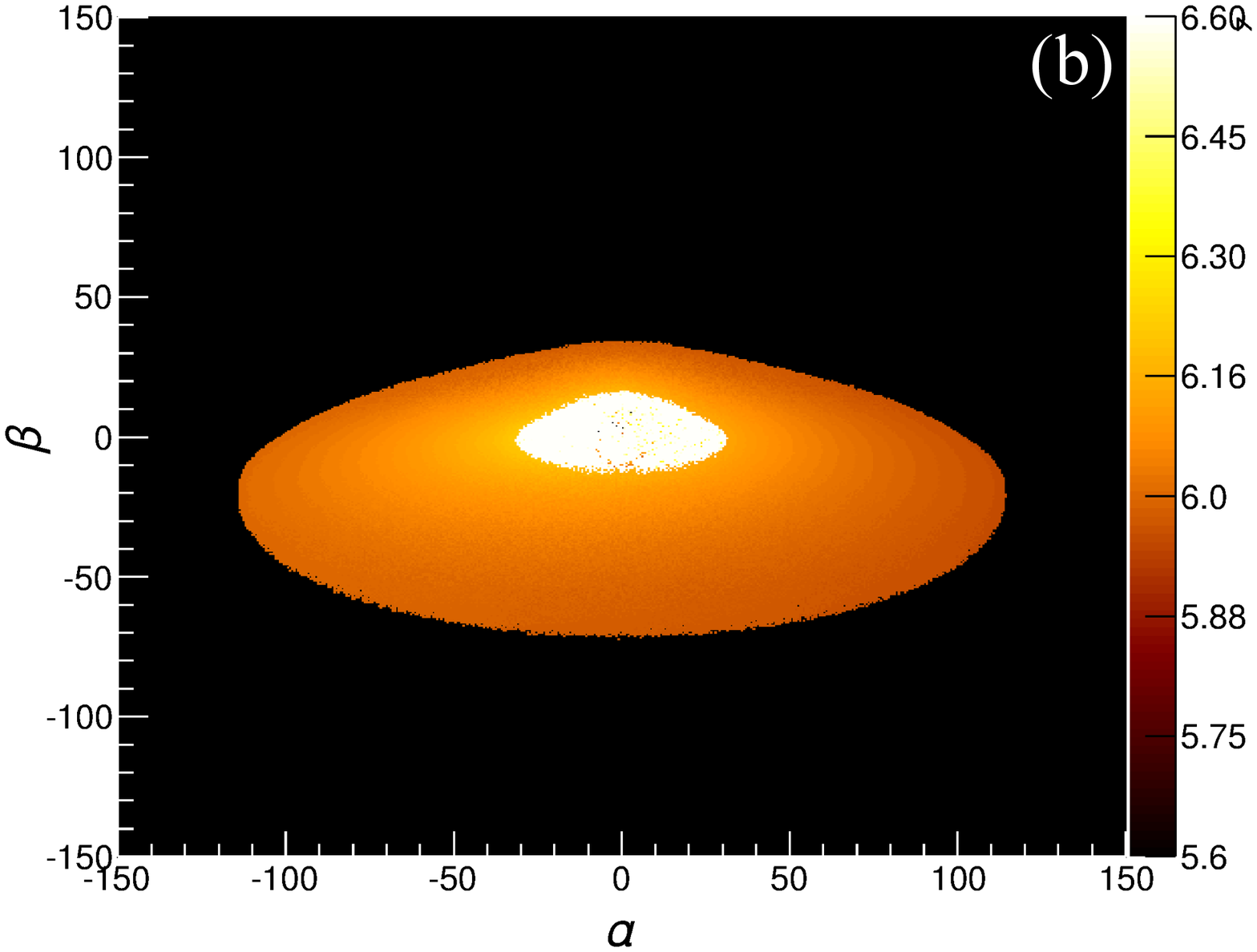}}
\caption{Images of TCAF (in $log(T_{obs})$ scale) as seen from the inclination angle of $50^\circ$, 
$70^\circ$ in presence of Compton cooling. Colorbar is in the range $10^{5.6}-10^{6.6}$ K. If the 
outflows are absent, dark central region of black hole just starts to appear to an observer at $50^\circ$ and 
becomes clearer if the onlooker moves to a lower inclination angle.}
\end{figure}

\vskip 0.5cm

\subsection{Realistic Observed Image with a finite beamwidth}

Figure 12, panel (a) shows a convolved image of a Keplerian disk with the inner edge 
at $3$ and outer edge at $50$. Convolution is over a beam width of (a) $\sim 20 r_g$ 
and (d) $\sim 5 r_g$.  Corresponding spectrum for this image gives a pure soft spectrum 
(see CT95 and SS73 for further details). 
The intensity variation due to the disk and the black hole can be seen. In Panels (b,e), 
the images show an accretion disk with 
CENBOL (same as Fig. 9(i) of CCG17a). 
 In this case, the spectrum is hard and the CENBOL surface completely screens the inner 
region close to the black hole. This case was studied in CCG17a. In Panel (c,f) 
outflows are strong as in an intermediate states and the convolutions show complex 
structure. If the inclination angle is very low, then possibly the black hole horizon 
could be seen, but in the present case of high inclination angle, the identification of
horizon would be difficult, especially if integrated over time. The images we just presented are
relevant for our galactic center under a sudden high mass inflow induced by 
tidal disruption. Various measurements (Markoff et al. 2007 and 
references therein) suggested that the jet from the Sgr A* is around $75^\circ$ away 
from our line of sight. Assuming this is along the perpendicular direction  with respect to the accretion plane, and
assuming radio intensity roughly tracks the disk/jet as discussed above, the Galactic centre
should also show images similar to the panels Fig. 12(a-f).
Our conclusion is that we may not be able to discern the horizon itself, but some 
features as in our Fig. 12(a-f) should be visible. Similarly the presence and absence of accretion disk 
with change in accretion rates should be seen.
     
\begin{figure*}
\centering
\vbox{
\includegraphics[width=2.0\columnwidth]{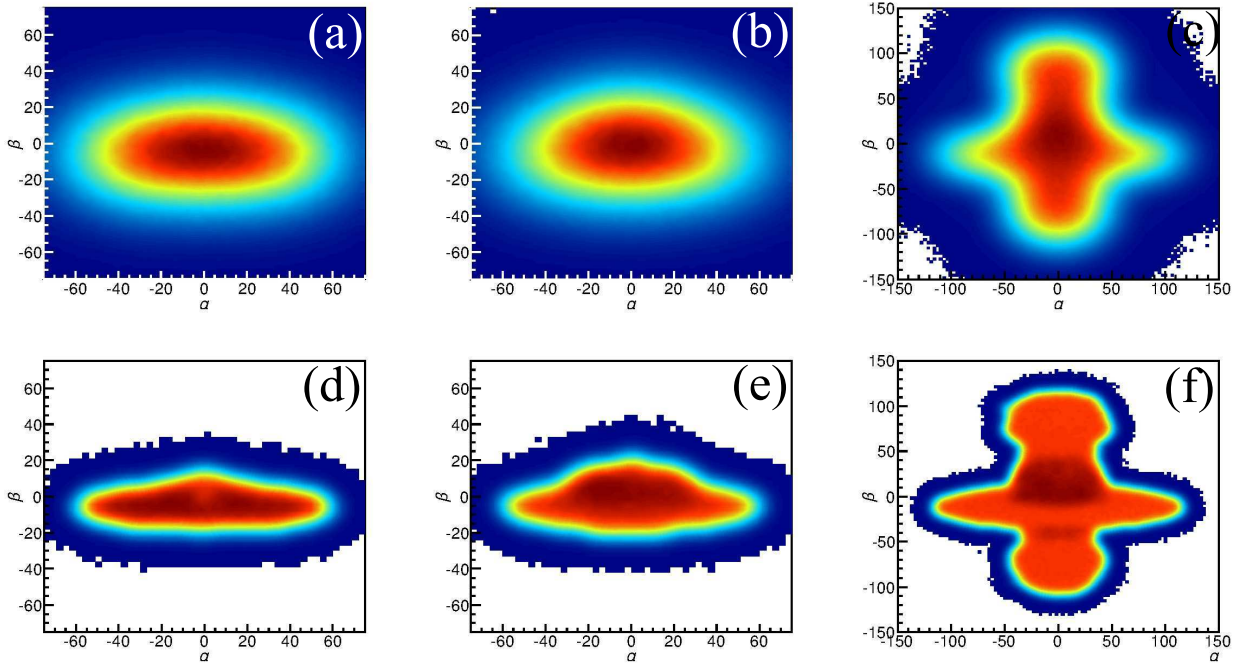}}
\caption{Convolved images of TCAF at different states. 2D Gaussian distribution is
considered for convolution with spreading diameter (hypothetical beam width of the instrument)  
of $20~r_g$ (for upper panel) and $5~r_g$ for lower
panel. Panels (a,d) are for a pure Keplerian disk (hard state with weak outflow) with outer edge 
at $50~r_g$. Panels (b,e) show 
images of TCAF without outflows with outer edge at $50~r_g$ (Soft state). Panels (c,f) depict images 
of TCAF with outflows having outer edges up to $100~r_g$. Inclination angle is $80^\circ$ for all 
panels. Intensity variation close to the black hole could be seen only when the beam width is around  
$5~r_g$.}  
\end{figure*}

The convolved images presented in Fig. 12 suggest a low spatial resolution of about $5~r_g$ 
will be able to distinguish features in the hole-disk system, unless the object has a very low 
inclination. Integrated counts from the central region is plotted in Fig. 13b for the Keplerian disk only. 
A spatial resolution of $5~r_g$ shows double-peaked curve in just resolved condition. Figure 13a shows
unresolved count distribution when $20 r_g$ is used instead. Considering the possibilities 
of high mass inflow onto Sgr A*, in its flaring state, we expect an X-Ray image similar 
to Fig. 8 which is difficult to resolve at a high inclination. Thus, 
probing inner region to capture the event horizon can be difficult in this state. Future projects 
which aim at capturing the image of an event horizon will not be able detect the difference in the intensity
of disk to that of the hole unless the object is in the soft state. Otherwise, only 
low inclination disks would show the even horizon in any spectral states, even in presence of 
outflows. Of course, this conclusion is valid only if the radio emission roughly tracks the inflow-outflow features as
observed in our numerical simulations.

\begin{figure}
\centering
\vbox{
\includegraphics[width=1.00\columnwidth]{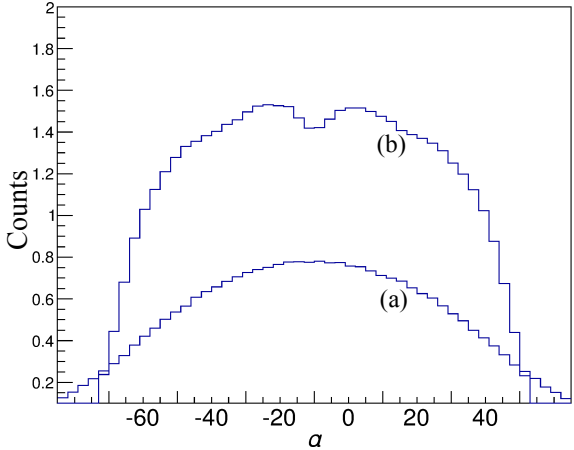}}
\caption{Difference in counts from central region received by detectors with spatial resolution of
$20~r_g$ (curve (a)) and of $5~r_g$ (curve (b)). This is case is for a pure Keplerian disk 
presented in Fig. 12(a,d).}
\end{figure}
\vskip 0.5cm

\section{Discussions and Conclusion}
Recent study of the effects of the photon bending on standard and two component 
accretion disk spectra and images was made by CCG17a. Earlier, static images of 
Two component advective flows were presented where we discussed how the Keplerian 
disk (valid for soft states only) and the two component disks with its intrinsic 
Compton cloud at the inner edge would look like from various angles. However, a 
realistic advective flow  will be time dependent and also will have outflows as 
the outflow and inflow solutions are always treated with a common footing (Chakrabarti, 
1989). This outflow must come out from the post-shock region or CENBOL which behaves 
as the Compton cloud around a black hole or a neutron star. This has also been 
shown by numerical simulations through numerous works referred to in the earlier 
Sections. The outflow could be time varying and the image of the disk would, in 
general, vary with time. 

In the present paper, we have employed numerical simulations of hydrodynamic process to 
generate time dependent accretion disk configurations. At every time step, Monte-Carlo 
simulation was done to compute the effects of Comptonization and to obtain the spectrum. 
We produced images of the CENBOL with the jet as well as the composite spectra 
when the effects of cooling due to Comptonization was turned off or turned on. We 
find that with cooling effects the CENBOL collapses to a smaller size and the outflow 
is also reduced as in a soft state. In presence of outflows, we draw images of an accretion 
disk where the Compton cloud or CENBOL is showing temporal variation of its physical 
shape and internal properties. With these, the images depict the outflows which are 
self-consistently produced from the inflowing matter. We showed the combined effects 
of the disk and the jet and particularly how they would look like in different 
X-ray energies. Doppler shift was seen to break the symmetry as expected, but 
the photon bending made it even more asymmetric. 

Though our simulations are carried out in regions which emit X-rays, and this 
will remain true for higher mass black holes also, but the efforts are on to 
observe the event horizon in radio waves which have higher spatial resolutions. 
On the other hand, both the radio and the Comptonization track the high energy 
electron distribution. So we believe that our simulations will have relevance for 
radio observations as well. However, intensity of synchrotron photon in 
radio band depends on the magnetic field strength and are expected to be primarily originated 
from the post shock region. Thus, our X-ray images without the cooler 
Keplerian disk, can be treated as radio images. We showed, using two different spatial resolutions, 
that one requires a resolution of $\sim 5 r_g$ in order to separate distinct features 
in disk-jet systems when high inclination objects are observed, unless the object 
is in a soft state and only has the Keplerian disk. However, the obscuration of the 
central region is not done when viewed at low inclination angle and the event horizon 
should be observable at any spectral state if the beam width is low enough. 
The images are of particular interest where the mass accretion rate onto central 
engine is high enough, such as RL Quasars, Seyfert 1 galaxies. The current
findings would hold for LLAGNs, like Sgr A*, for high mass inflow induced by tidal 
disruptions. Considering the future possibilities of observing such Galactic black holes and SMBHs, we 
present an example of a high inclination system, to impress how the inclination 
basically removes the chances of observing the horizon, especially, when the 
jets are active. In the absence of jets and outflows, however, the horizon can be 
discerned. This result will be of importance for future missions which could exclude
high inclination Seyfert 2 AGNs and GBHs such as GRS 1915+105, H 1743-322, GRO 1655-40. 

In the spectral study using Monte Carlo simulations, we see effects of 
photon bending in the multi-color soft photon component. The peak was seen at a lower 
energy due to redshifts at an inclination of $70^{\circ}$. We also find that the 
spectra harden at higher inclination angles. Of course, a major result, though 
expected, in our simulation is that when the cooling is included, the CENBOL itself shrinks 
and consequently the spectra become softer. Further simulations including the effects 
of synchrotron radiation, Compton cooling and viscosity are being carried 
out and will be reported elsewhere. 
 
\section{Acknowledgements}
This research was possible in part due to a grant from Ministry of Earth Sciences 
with Indian Centre for Space Physics.



\end{document}